\documentclass[preprint]{aastex}
\usepackage{graphicx}
\usepackage{amsmath}

\def\Soh{{\Sigma_{_0}}}
\def\hoh{{h_{_{0}}}}
\def\aoh{{a_{_{0}}}}
\def\ain{{a_{_{\rm in}}}}
\def\aindot{{\dot{a}_{_{\rm in}}}}
\def\aout{{a_{_{\rm out}}}}
\def\apo{{r_{_{\rm\!apo}}}}
\def\peri{{r_{_{\rm\!peri}}}}
\def\orchestra{\textit{Orchestra}}

\newcommand{\lae}{\lower 2pt \hbox{$\, \buildrel {\scriptstyle <}\over {\scriptstyle\sim}\,$}}
\newcommand{\gae}{\lower 2pt \hbox{$\, \buildrel {\scriptstyle >}\over {\scriptstyle\sim}\,$}}

\def\eqnum[#1]{(\ref{#1})}

\def\Msolar{{\rm M}_\odot}
\def\Mearth{{\rm M}_\oplus}
\def\Mjupiter{{\rm M}_J}
\def\mearth{\ifmmode {\rm M_{\oplus}}\else $\rm M_{\oplus}$\fi}
\def\msun{\ifmmode {\rm M_{\odot}}\else $\rm M_{\odot}$\fi}

\def\vecdv{\vec{v}_{_{\rm rel}}}

\def\rp{r_\bullet}

\def\Cdyn{{C}_{_{\rm df}}}

\def\mp{m_\bullet}
\def\rhop{\rho_\bullet}
\def\rhog{\rho_{_{\rm gas}}}
\def\cs{c_s}
\def\Rsonic{R_s}

\def\ma{\mu}

\def\Mdisk{M_{_{\rm disk}}}

\def\vkep{v_{_{\rm Kep}}}

\def\openrate{{\kappa_{_{\rm in}}}}

\def\PlanetX{Planet~Nine}

\def\biden{2012~VP$_{113}$}

\begin{document}

\title{Making Planet Nine: A Scattered Giant in the Outer Solar System}

\author{Benjamin C. Bromley}
\affil{Department of Physics \& Astronomy, University of Utah, 
\\ 115 S 1400 E, Rm 201, Salt Lake City, UT 84112}
\email{bromley@physics.utah.edu}

\author{Scott J. Kenyon}
\affil{Smithsonian Astrophysical Observatory,
\\ 60 Garden St., Cambridge, MA 02138}
\email{skenyon@cfa.harvard.edu}

\begin{abstract}

Correlations in the orbits of several minor planets in the outer solar
system suggest the presence of a remote, massive Planet Nine. With at
least ten times the mass of the Earth and a perihelion well beyond
100~AU, Planet Nine poses a challenge to planet formation theory.
Here we expand on a scenario in which the planet formed closer to the
Sun and was gravitationally scattered by Jupiter or Saturn onto a very
eccentric orbit in an extended gaseous disk. Dynamical friction with
the gas then allowed the planet to settle in the outer solar system.
We explore this possibility with a set of numerical simulations. 
Depending on how the gas disk evolves, scattered super-Earths or 
small gas giants settle on a range of orbits, with perihelion distances 
as large as 300~AU. Massive disks that clear from the inside
out on million-year time scales yield orbits that allow a super-Earth
or gas giant to shepherd the minor planets as observed.  A massive
planet can achieve a similar orbit in a persistent, low-mass disk over
the lifetime of the solar system.

\end{abstract}


\maketitle

\section{Introduction}

The orbital alignment of minor planets located well beyond Neptune,
including Sedna and \biden, inspired \citet{trujillo2014} to invoke a
massive, unseen planet orbiting at roughly 200~AU from the Sun.
Expanding on this analysis, \citeauthor{batygin2016} (\citeyear{batygin2016};
see also
  \citealt{brown2016}) propose a more distant planet
which maintains the apsidal alignment for a set of six trans-Neptunian
objects. With a mass more than ten times that of the Earth, this
planet would have a semimajor axis between 300~AU to 1500~AU, an
eccentricity within the range of roughly 0.2--0.8, an inclination
below 40$^\circ$, and an apsis that is anti-aligned with the six minor
planets. Subsequent work by \citet{fienga2016}, using precise Cassini
radio ranging data of Saturn, places constraints on the perturber's
orbital phase.

The prospect of a \PlanetX\ lurking in the outer solar system provides 
a new opportunity to test our understanding of planet formation theory.
Various mechanisms -- coagulation \citep{kb2015a}, gravitational 
instability \citep[][and references therein]{helled2014}, and 
scattering \citep{bk2014} -- can 
place a massive planet far from the host star. Aside from a direct
detection of \PlanetX, testing these ideas requires numerical 
simulations which predict the properties of planets as a function of
initial conditions in the protoplanetary disk.

Previous calculations of gas giant planet formation \citep{rasio1996,
  weiden1996, ford2005, moorhead2005, lev2007, chatterjee2008,
  bk2011a} demonstrate that growing gas giants clear their orbital
domains by scattering super-Earths or more massive planets to large
distances.  If the surface density of the gaseous disk at large
distances is small, scattered planets are eventually ejected. For
disks with larger surface densities, however, dynamical friction damps
a scattered planet to lower eccentricity \citep{dokuchaev1964,
  rephaeli1980, takeda1988, ostriker1999,
  kominami2002}. \citet{bk2014} used simple models of disk-planet
interactions to show that this mechanism plausibly circularizes the
orbits of massive planets at 100--200~AU from the central star.

The \citet{batygin2016} analysis poses a new challenge to scattering
models. Although they propose a moderately eccentric orbit for their
massive perturber, the planet has a semimajor axis
beyond 300--400~AU.  Our goal here is to show under what conditions a
scattered planet can achieve this orbit through dynamical friction
with a gas disk. We consider a wide range of planet and disk
configurations, numerically simulate outcomes of these models, and
assess how well they explain a \PlanetX\ in the outer solar system.

\section{Method}

To explore the possibility of a scattered origin for \PlanetX, we
follow our earlier strategy \citep{bk2014}.  We choose initial 
conditions for planetary orbits and gas disks and a mechanism for 
disk dissipation. We then track the orbital evolution with the 
$n$-body integration component of our \orchestra\ code
\citep{bk2006, kb2008, bk2011a}. In this section we provide 
details of the disk models and an overview of the numerical method, 
which includes an updated treatment of dynamical friction.

\subsection{The disk models}

To set the stage for relocating a scattered planet from 
5--15~AU into the outer solar system, we model
the Sun's gas disk with the following prescription
for surface density $\Sigma$, scale height $H$, and 
midplane mass density $\rhog$:
\begin{eqnarray}
\label{eq:S}
&\ &
\Sigma(a,t)  =  
\begin{cases} 
\Soh 
\left(\frac{a}{\aoh}\right)^{\!-1}  
e^{-t/\tau}, & \textrm{ if } \ain \leq a \leq \aout, \\
0 & \textrm{ otherwise,}
\end{cases}
\\
&\ &
H(a)  = \hoh a \left(\frac{a}{\aoh}\right)^{\!\,2/7},
\\[7.5pt]
&\ & 
\rhog(a,t) =  \frac{\Sigma(a,t)}{H(a)} ~ .
\end{eqnarray}
Here, $\Soh$ sets the surface density at distance $\aoh \equiv 1$~AU,
$\ain$ and $\aout$ are the inner and outer edges of the disk, and
$\hoh = 0.05$ establishes the scale height of the flared disk
\citep{kh1987, chiang1997, and2007, and2009}. The global surface 
density decay parameter $\tau = 1$--10~Myr enables a homologous 
reduction in the surface density \citep{haisch2001}. To allow the 
inner edge of the disk to expand, 
as in a transition disk, 
we adopt an expansion rate 
$\openrate$:
\begin{equation}
  \ain(t) = \ain(0) + \openrate t ,
\end{equation}
where the initial size of the inner cavity is $\ain(0)\equiv 20$~AU.
Observations of transition disks \citep{calvet2005, currie2008,
  and2011, najita2015} suggest opening rates of O(10)~AU/Myr. 


Toward estimating dynamical friction and gas drag, we assume that the
sound speed in the gas is $\cs \approx H\vkep/a$, where $\vkep$ is the
circular Keplerian speed at orbital distance $a$ from the Sun.  We
also assume that both $H$ and $\cs$ are independent of time. Armed
with these variables we can determine the Mach number of a planet
moving relative to the gas, and hence derive drag forces on the
planet.  These estimates include the effect of pressure support within
the gas disk, which makes the bulk flow in the disk sub-Keplerian,
with an orbital speed that is reduced from $\vkep$ by a factor of
$(1-H^2/a^2)$ \citep[e.g.,][]{ada76,weiden1977a,yk2013}.

Table~\ref{tab:parmsdisk} lists disk model parameters. We distinguish
two types of disks: static and evolving. Static disks have fixed
surface density profiles and small total mass, $0.002 \Msolar < \Mdisk
< 0.06 \Msolar$ (2--60~$\Mjupiter$, where $\Mjupiter$ is the mass of Jupiter), 
which extend to 1600~AU.  These models enable us to consider the 
possibility of long-term ($\gtrsim$100~Myr) planet-disk interactions. 
Evolving disks extend to 800~AU with larger initial surface density 
and mass.  In the most extreme case ($\Soh = 1000$~g/cm$^2$), the 
disk mass is half that of the Sun. 
Although improbable \citep[cf.][]{and2013}, this extreme disk mass
allows us to explore the possibility of rapid, strong orbital damping.
Because dynamical friction depends on the gas density, $\rhog\sim
\Sigma/H$, we can scale results to less massive disks with smaller
scale heights.

\subsection{Scattered planets}

For each disk configuration in Table~\ref{tab:parmsdisk}, we carry out
simulations with planets scattered to large ($> 1000$~AU) distances.
As summarized in Table~\ref{tab:parmsplanet}, each planet is assigned 
a mass $\mp$ and mean density $\rhop$, from which we infer a physical 
radius, $\rp$. In our orbital dynamics code, the planet is launched 
from a perihelion distance of $\peri = 10$~AU with a speed that would 
take it to a specified aphelion distance $\apo$ if it were
on a Keplerian orbit about the Sun.  We track the subsequent dynamical
evolution with orbital elements calculated geometrically, since the
disk potential can complicate the interpretation of osculating orbital
elements.  While the planet's orbit typically comes close to the nominal 
starting aphelion, it never makes it to $\apo$ exactly due to dynamical 
friction with the gas and the disk's overall gravitational potential.

\subsection{Numerical approach}

To evolve planetary orbits in a gas disk, we follow \citet{bk2014}. 
We use the orbit integrator in our hybrid $n$-body--coagulation code 
\orchestra\ to calculate the trajectory of individual planets around 
the Sun in the midplane of the disk. We calculate disk gravity using 
2000 radial bins spanning the planet's orbit, assigning a mass to 
each bin according to Equation~(\ref{eq:S}). We initially solve the 
Poisson equation by numerical integration, storing the results. The 
saved potential is updated as the disk evolves. 

To estimate acceleration from dynamical friction, we adopt a
parameterization similar to \citet{lee2014} in the absence of 
gas accretion \citep[see also][]{dokuchaev1964, ruderman1971, ostriker1999}:
\begin{equation}\label{eq:drag}
  \frac{d\vec{v}}{dt}  =  
    -\frac{G^2 \rhog \mp \ma}{\cs^2} 
    \frac{(1+4\pi^2\Cdyn^2\ma^2)^{1/2}}{(1+\ma^2)^{2}} \frac{\vecdv}{|\vecdv|},
\end{equation}
where $\vecdv$ is the planet's velocity relative to the gas,
$\ma \equiv |\vecdv|/\cs$ is the mach number, and $\Cdyn$ is
a constant that depends on the geometry of the disk in the
plane perpendicular to the planet's motion.

In evaluating the coefficient $\Cdyn$, we previously only considered 
contributions from gas more distant than $H/2$ from a planet \citep{bk2014}. 
Here we are less restrictive and include contributions from material 
closer to the planet.  The drag acceleration thus has a piece from 
distant disk material in slab geometry \citep{bk2014}, along with a 
Coulomb logarithm \citep[e.g.,][]{binn2008}:
\begin{equation}
\Cdyn \approx 0.31 + \ln\left[\max(1,\frac{H}{2R})\right],
\end{equation}
where the radius $R$ is 
\begin{equation}
R \equiv \max(\rp, \Rsonic)
\end{equation}
and 
\begin{equation}
\Rsonic
 = \frac{1}{\ma_x^2-1} \frac{2 G\mp}{cs^2} 
\ \ \ \ \ [\ma_x^2 = \max(\ma^2,1.0001)]
\end{equation}
is an effective sonic radius \citep[e.g.,][]{thun2016}. 

With this formulation, our goal is to map how scattered planets with
large eccentricities ($e\lesssim 1$) damp to modest values ($e \sim
0.5$) when the planet moves at supersonic speeds. Once the planet
achieves low eccentricity, its subsequent evolution is complicated by
differential torque exchange with the disk \citep{gold1980, ward1997}
and accretion of disk material \citep{hoyle1939, lee2011}.  We do not
attempt to track this behavior. Our prescription underestimates
dynamical friction in the subsonic and transonic regimes
\citep[cf.][]{ostriker1999}. Thus we follow a planet as its orbit
damps, but stop the integration if it manages to fully circularize
before the gas disappears.

\section{Results}

We ran over 10$^4$ simulations to map out the parameter space of disk
and planet configurations. To describe our results, we first consider 
low-mass, static disks, which isolate the physics of dynamical damping 
without the complications of disk evolution. To evaluate damping outcomes
over the 1--10~Myr lifetimes of typical protoplanetary disks, we then 
consider a set of evolving disks.

\subsection{Relocation of a scattered planet in a long-lived disk}

\begin{figure}[htb]
\centerline{\includegraphics[width=7.0in]{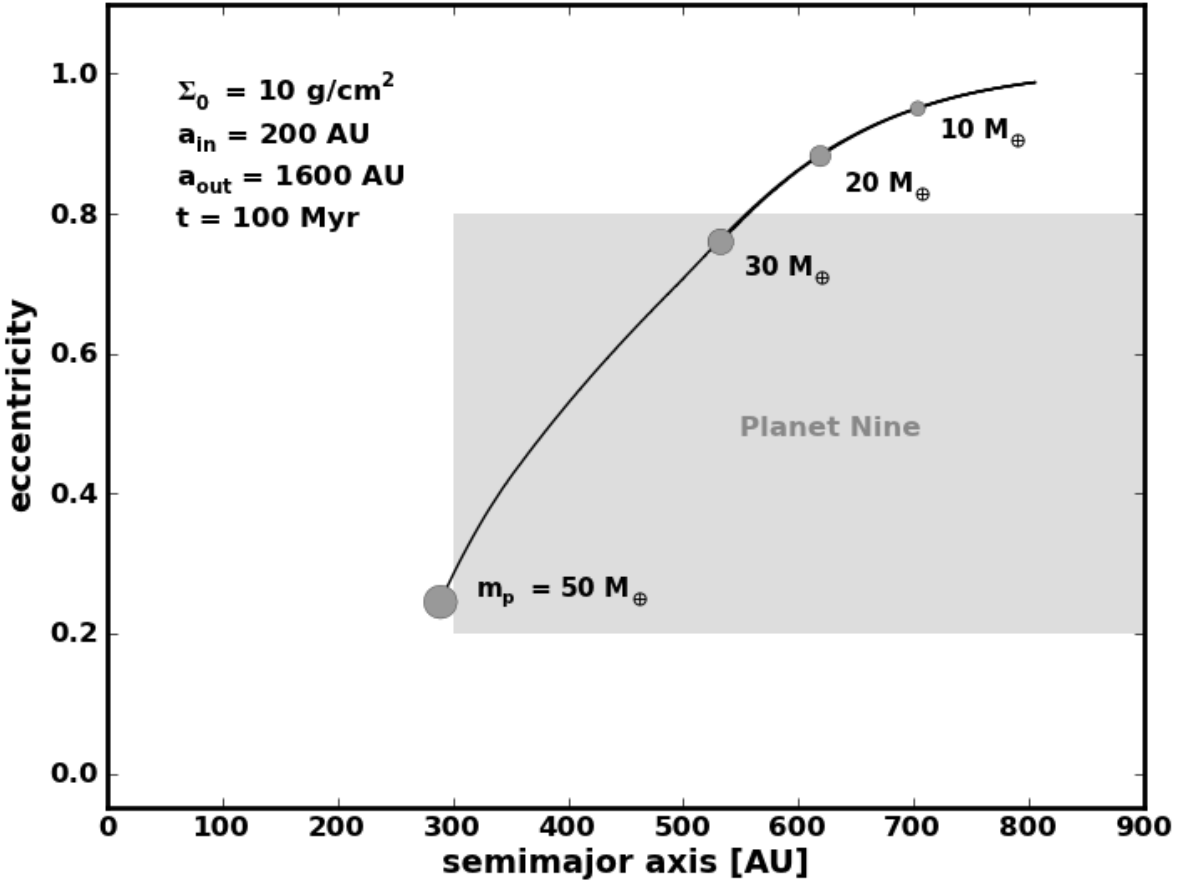}}
\caption{\label{fig:aetm} Simulations of the orbital evolution of
  scattered planets in a static gas disk. Solid curves show the
  semimajor axis ($a$) and eccentricity ($e$) of four planets with
  various masses as they evolve over a period of 100~Myr. The 
  legend indicates the set of disk parameters.  All planets start 
  at the same high $a = 800$~AU and $e = 0.9875$ and evolve along a 
  single path towards smaller values of $a$ and $e$.  circular 
  symbols show the final outcomes after 100~Myr, labeled with planet 
  mass. More massive planets evolve faster and progress further along 
  the path.
%
  The gray region approximates the allowed range of $a$ and $e$ 
  from \citet{batygin2016} for \PlanetX.}
\end{figure}

In this set of simulations, we set up static disks with low surface density
($\Soh = 2$--50~g/cm$^2$), large radial extent ($\aout = 1600$~AU),
and big inner cavities ($\ain = 50$--200~AU). We evolve scattered
planets over a time $t = 100$~Myr. Figure~\ref{fig:aetm} shows several
outcomes with planets of different masses.  All planets follow the 
same track in $a$--$e$ space as they circularize; more massive planets
evolve further along the track.

\begin{figure}[htb]
\centerline{\includegraphics[width=7.0in]{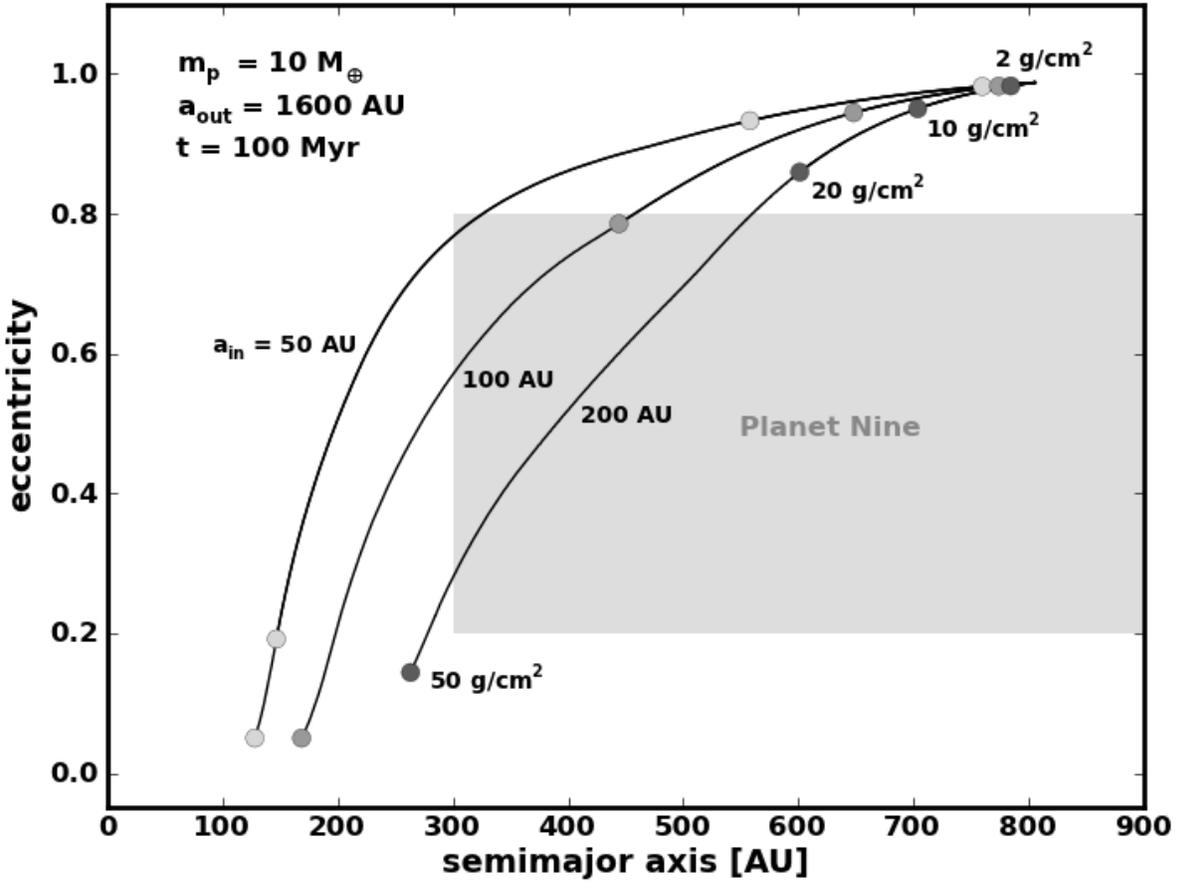}}
\caption{\label{fig:aetd} Orbital evolution of a planet in various 
  configurations for a static disk. As in Fig.~\ref{fig:aetm}, planets
  starting with large $a$ and $e$ follow specific tracks which depend
  on the inner edge of the disk ($\ain$). Larger inner cavities allow
  planets to settle at larger $a$.  For a fixed planet mass of 
  10~$\Mearth$, the final location in the $a$--$e$ plane depends on 
  the surface density parameter ($\Soh$).  In disks with higher surface 
  density, orbits evolve more quickly and reach smaller $a$ and $e$. }
\end{figure}

Figure~\ref{fig:aetd} illustrates how evolution depends on the disk
configuration. The plot shows three separate evolutionary tracks in
$a$--$e$ space, each corresponding to a different inner edge of the
disk ($\ain$). With a smaller inner edge, the disk causes a planet
to settle more quickly (because there is more disk material to
interact with) and closer to the Sun. The markers in the plot
designate how far each planet evolves. Planets in low surface density 
disks make less progress along their track than those in high surface 
density disks. 

These calculations establish an approximate degeneracy between mass
and the surface density parameter $\Soh$ in the formula for dynamical
friction acceleration (Equation~(\ref{eq:drag})). As a result, the
progress that a planet makes along its $a$--$e$ track in a fixed
amount of time depends only on the product, $\mp\times\Soh$.  Thus,
data points showing the orbital evolution as a function of planet mass
in Figure \ref{fig:aetm} can also represent the progress of a planet
of fixed mass in disks with different surface densities.  Similarly
points in Figure \ref{fig:aetd} can represent outcomes with different
planet masses at fixed surface density.

If long-lived disks are responsible for settling a planet on the type of 
orbit inferred by \citet{batygin2016}, then our suite of simulations 
suggests the following condition leads to successful \PlanetX-like outcomes:
\begin{equation}
\left(\frac{\Soh}{10~\textrm{g/cm$^2$}}\right)
\left(\frac{\mp}{20~\Mearth}\right) \approx
\left(\frac{\ain}{100~\textrm{AU}}\right)^{1/2}
\left(\frac{t}{100~\textrm{Myr}}\right)^{-1} ~ .
\end{equation}
This expression applies as long as $\aout \gg \ain$.
While it is only approximate, this relation suggests that a
persistent, low-mass disk ($\Soh \approx 0.3$~g/cm$^2$; about a
quarter of a Jupiter mass in gas) can modestly damp a scattered
Neptune-size body within the age of the solar system. 

\subsection{Settling in an evolving disk}

Observations indicate that the youngest stars are surrounded with
opaque disks of gas and dust \citep[see][]{kh1995, kgw2008, will2011,
  and2015}.  Surface densities vary; the ``Minimum Mass Solar Nebula'' 
value of $\Soh \approx$ 2000~g/cm$^2$ \citep{weiden1977b,hayashi1981} 
is at the upper end of the range observed in the youngest stars 
\citep[e.g.,][]{and2013, najita2014}.
These disks globally dissipate on a time scale, $\tau$, of millions of
years \citep{haisch2001}, and may also erode from the inside out at
a rate of $\openrate \gtrsim O(10)$~AU/Myr, as in transition disks
\citep[e.g.,][]{and2011, najita2015}. The resulting behavior of a
scattered planet as it settles depends sensitively on how mass is
distributed in these disks as they evolve.  

\begin{figure}[htb]
\centerline{\includegraphics[width=7.0in]{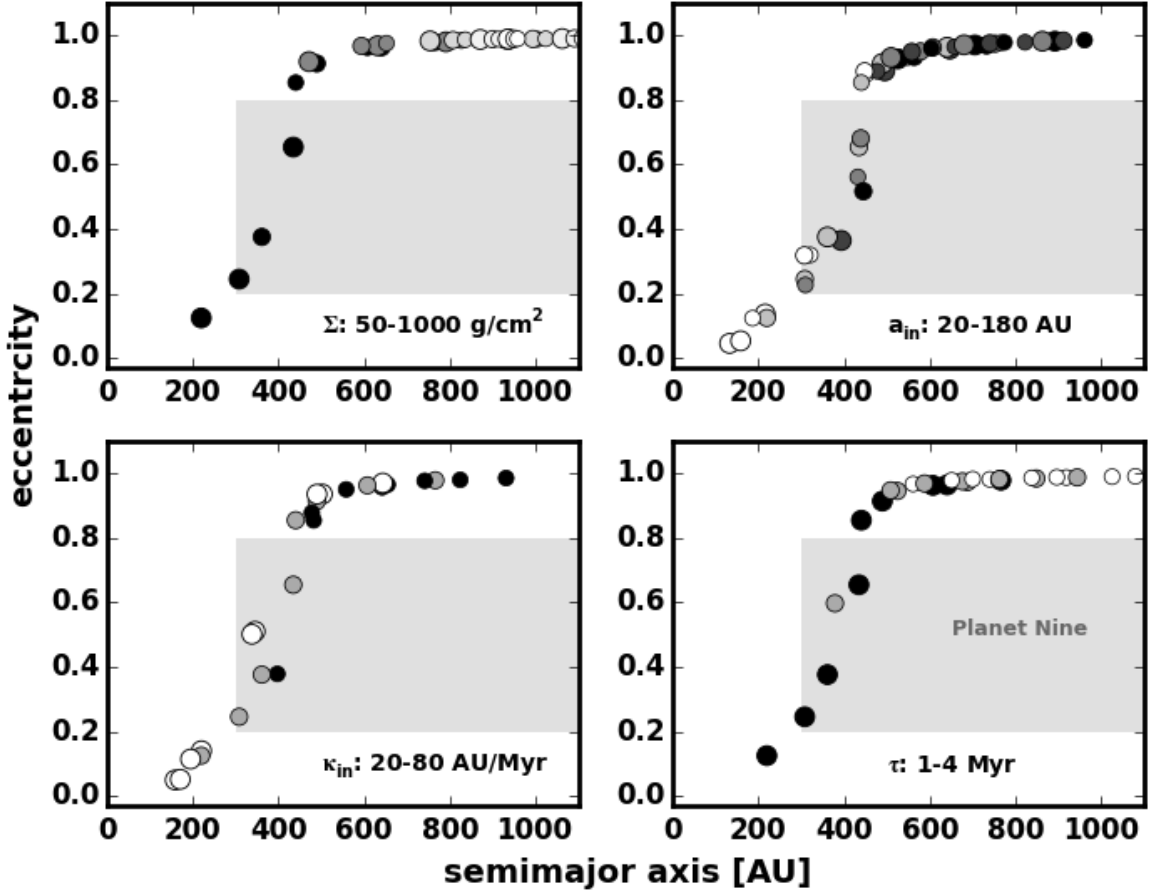}}
\caption{\label{fig:aem} 
  Semimajor axis and eccentricity at 10~Myr for scattered planets 
  with masses between 15~$\Mearth$ and 30~$\Mearth$ in evolving disks
  with baseline parameters $(\Soh,\ain,\openrate,\tau)$  = 
  (1000~g/cm$^2$, 60~AU, 40~AU/Myr, 4~Myr) and a range of initial 
  aphelion distances (2000--2800). Each panel illustrates how outcomes 
  change when one of these parameters is varied in the range specified 
  in the legend. Symbol shades indicate the value of the varied parameter 
  from white (lower values) to black (upper values).  In all panels, 
  symbol size correlates with planet mass.  }
\end{figure}

Figure~\ref{fig:aem} illustrates the dependence of planetary settling
on disk parameters $\tau$, $\openrate$, the initial disk surface density,
and the inner disk edge. Adopting a baseline model where 
($\Soh$,$\ain$,$\openrate$,$\tau$) = (1000~g/cm$^2$, 60~AU, 40~AU/Myr, 4~Myr), 
we vary individual disk parameters for planets with masses of 15--30~$\Mearth$, 
scattered to starting distances of $\apo = 2000$--2800~AU. The general trends 
are clear. Dynamical settling to small orbital distance and low eccentricity 
is more effective in massive, slowly evolving disks with small inner cavities.  

Figure~\ref{fig:aep} summarizes the outcomes of all of our evolving
disk models (see Tables~\ref{tab:parmsdisk} and
\ref{tab:parmsplanet}). The trends that emerged in
Figure~\ref{fig:aem} are apparent in this Figure as well: long-lived,
massive disks lead to significant dynamical evolution, while
short-lived, low-mass disks do not.  Figure~\ref{fig:aep} also shows
an extended ``sweet spot'' in $a$--$e$ space, labeled with ``Planet
Nine,'' roughly corresponding to orbital elements of the massive
perturber hypothesized by \citet{batygin2016} and \citet{brown2016}.
Several hundred models yield planets that lie in the sweet spot,
suggesting that the scattering mechanism can explain the inferred
orbit of \PlanetX\ in the outer solar system.

Despite the trends revealed in Figures~\ref{fig:aetm}--\ref{fig:aem},
it is difficult to tell which set of model parameters leads to
successful \PlanetX-like orbits. To distinguish models in a way
that highlights successful ones, we define two variables,
\begin{eqnarray}
\label{eq:P}
P & \equiv & \left(\frac{\mp}{10~\Mearth}\right)
\left( \frac{\Soh}{1000~\textrm{g/cm}^2}\right)
\left(\frac{\apo}{1000~\textrm{AU}}\right)^{-1}
\\
\label{eq:Q}
Q & \equiv & 
\left(\frac{\tau}{1~\textrm{Myr}}\right) 
\left[\left(\frac{\openrate}{60~\textrm{AU/Myr}}\right)
\left(\frac{\ain}{\textrm{20 AU}}\right)
\left(\frac{\apo}{\textrm{1000 AU}}\right)\right]^{-1}.
\end{eqnarray}
Roughly, the first variable is a mass-dependent damping rate,
determined by planet mass and the disk mass, along with a factor of
$1/\apo$ that reduces this rate if the planet is launched further away
from the Sun.  The second one measures the disk lifetime, based on 
the global disk decay time and the time for the disk to clear
from the inside out, along with geometric factors involving the disk's
radial extent and the planet's initial orbit.  For models where
$\tau$ is formally infinite, we set $\tau = 10$~Myr, the simulated
duration of the evolving disk models.

The variables $P$ and $Q$ help to isolate the parameters that are
necessary for a scattered giant planet to settle on a \PlanetX-like orbit.
Our choice for defining these quantities, along with the mass, length
and time scales in Equations~(\ref{eq:P}) and (\ref{eq:Q}), are based
more on simplicity than anything else; other combinations of model
parameters may serve the same purpose. Nonetheless, our choice 
yields a nicely compact region in $P$--$Q$ space for models that
succeed in matching Batygin and Brown's (2016) criteria for \PlanetX.

Figure~\ref{fig:pixx} shows a swath of points in the $P$--$Q$ plane
that correspond to successful models. These points have just the right
balance between the masses of the disk and the planet ($P$) on the one
hand, and disk lifetime ($Q$) on the other.  Models without this
balance tend to produce planets that 
circularize at small semimajor axes (high $P$ and high
$Q$; large masses and long disk lifetimes) or remain highly eccentric
at large semimajor axes (low $P$ and low $Q$; small masses and short
disk lifetimes).

A rough quantitative relationship between $P$ and $Q$ for successful
models is $Q \sim P^{-3/2}$; in terms of model parameters, this
condition translates to:
\begin{equation}\label{eq:PQ}
\left(\frac{\mp}{1~\Mearth}\right)
\left(\frac{\Soh}{1000~\textrm{g/cm}^2}\right)
\approx
\left(\frac{\openrate}{40~\textrm{AU/Myr}}\right)^{\!2/3}
\!
\left(\frac{\ain}{\textrm{40 AU}}\right)^{\!2/3}
\!
\left(\frac{\tau}{1~\textrm{Myr}}\right)^{\!-2/3}
\!
\left(\frac{\apo}{1000~\textrm{AU}}\right)^{\!-1/3},
\end{equation}  
from which the inverse relationship between disk lifetime and mass
factors is apparent, as is the sensitivity to disk and orbit
geometries.

\subsection{Summary of simulation outcomes}

These simulations suggest a broad range of outcomes in $a$--$e$ space
for 1--50~\mearth\ planets scattered from 10~AU into the outer part of
a gaseous disk. For many combinations of input parameters, planets
remain on $e \gtrsim$ 0.80 orbits at $a \gtrsim$ 400~AU.  Another set
of parameters yields in massive planets on nearly circular orbits at
100--200~AU from the host star.  Specific combinations of disk and
planet properties result in ``successful models,'' with planets on
orbits consistent with the massive perturber of \citet{batygin2016}:

\begin{enumerate}
\item A planet scattered at low inclination into a low-mass,
  long-lived disk damps at a rate proportional to the planet mass and
  the disk's surface density. The final semimajor axis depends on
  $\ain$, the inner radius of the disk; successful models have $\ain
  \gtrsim 50$~AU.  A power-law disk ($\Sigma \sim 1/a$) with low
  surface density ($\Sigma \sim 3\times 10^{-3}$~g/cm$^2$ at 100 AU,
  extending to $\sim 1000$~AU) can produce a Neptune-size \PlanetX\ on
  a moderately eccentric orbit within the lifetime of the solar
  system. Higher planet masses or larger surface densities lead to
  success in less time.

\item Scattering in a massive, short-lived disk leads to \PlanetX-like
  orbits when there is a balance between the damping rate and the disk
  evolution time scale. In successful models, planet masses are
  typically 10~$\Mearth$ or more, although 5~$\Mearth$ planets can
  acquire a \PlanetX-like orbit in the most massive, long-lived disks.
  Most successful models experience either slow global decay ($\tau =
  4$~Myr) or none at all. The disk then evolves primarily through
  inside-out erosion, as in a transition disk. This feature helps
  successful planets settle at large semimajor axes.

\item In all of our simulations, successful models tend to have
  semimajor axes that lie within $a = 600$~AU. Batygin and Brown's
  preferred model has a higher orbital distance, with $a \approx
  700$~AU \citep[see also][]{molhatra2016, brown2016}. While their
  analysis accommodates a wide range of possibilities, our current
  models do not. If a massive perturber were to have a semimajor axis
  firmly established beyond 700~AU, our mechanism would require a disk
  with more mass beyond a few hundred AU.

\end{enumerate}

\section{Discussion}
\label{sect:discuss}


Although a \PlanetX\ in the outer solar system has not yet been
confirmed, several massive exoplanets have been identified at large
distances from their host stars. The outermost planet in the HR~8799
system has a semimajor axis of $a \approx$ 70~AU \citep{marois2008,
  maire2015}. The planets in 1RXS J160929.1$-$210524 \citep[$a
  \approx$ 330~AU;][]{lafren2010}, and HD~106906~b \citep[$a \approx$
  650~AU][]{bailey2014} have much larger semimajor axes.
Gravitational instability is a popular mechanism to produce planets
with such large $a$ \citep[e.g.,][]{helled2014, rice2016}. Our
calculations demonstrate that a planet scattered from $a \approx$
10~AU can interact with a gaseous disk and settle on roughly circular
orbits at much larger $a$. Thus, scattering is a viable alternative to
disk instability for placing massive planets at large $a$.

In our approach, we do not consider whether a scattered planet might
accrete gas as its orbit damps \citep[cf.][]{hoyle1939}. In principle,
planets at large $a$ might accumulate significant amounts of gas in
1--10~Myr. Whether accreting planets end up in configurations similar
to those of the gas giants in HR~8799, 1RXS J160929.1$-$210524, and
HD~106906~b requires an expanded set of more physically realistic
simulations which are beyond the scope of the present work.

Here, we have focused on identifying initial conditions 
that
yield a 
planet of fixed mass 
on an orbit with $e \sim$
0.2--0.8 at $a \gtrsim$ 300~AU.  With over $10^4$ models, we survey a
variety of planet masses, scattered orbits, and configurations of the
gas disk.  A large central cavity in the disk, as observed in some
transition disks \citep[e.g.,][]{and2011}, is essential to settling
\PlanetX\ at large orbital distances. Throughout, we assume that
scattering and subsequent damping occur at low inclination;
this condition is necessary for optimal interaction with the disk. A
low scattering inclination is also expected. Damping by gas and
planetesimals in the gas giant region likely kept larger bodies on
orbits that were nearly coplanar with the gas disk
\citep[e.g.,][]{liss1993b}.

Our ``successful'' models --- with outcomes that have the orbital
characteristics of Batygin and Brown's (\citeyear{batygin2016})
inferred massive perturber --- are those that balance planet and disk
masses with disk longevity.  In models where the disk is long-lived
but low-mass, a planet like Neptune can settle within a few billion
years.  Successful models with more rapid gas dissipation require more
massive disks. Disks that evolve on time scales of a few million years
can lead to \PlanetX-like orbits, only if the initial disk mass is
about 0.1~$\Msolar$ or more. A smaller disk scale height and/or
reduced flaring of the disk \citep[e.g.][]{keane2014} can reduce this
restriction on the disk mass.


In addition to our proposal for scattering and damping as an origin
for a massive perturber in the outer solar system, there are other
compelling possibilities. These include \textit{in situ} formation,
late-time dynamical instabilities (the Nice model), passing stars, and
Galactic tides. Each of these phenomena lead to different outcomes for
\PlanetX.

\begin{figure}[htb]
\centerline{\includegraphics[width=7.0in]{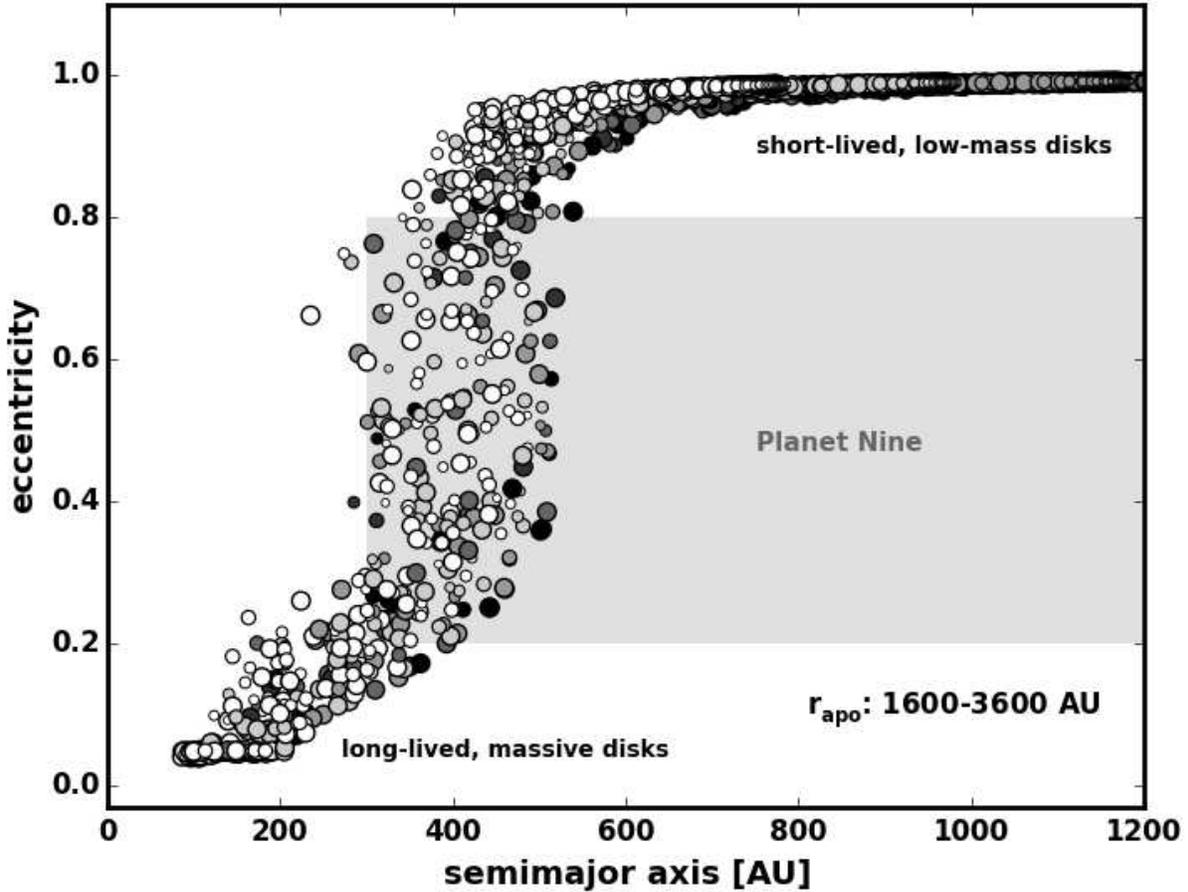}}
\caption{\label{fig:aep} 
  Outcomes for scattered planets at 10~Myr in evolving disks.
  Symbol size correlates with planet mass; shading indicates 
  initial aphelion distance (lightest: 1600~AU; darkest: 3600~AU). 
  More massive planets starting at the smallest aphelion distance 
  settle at smaller orbital distances with lower eccentricities. 
  Variations in the initial disk configuration and the mode/time scale
  for disk dissipation move the final $(a,e)$ along the sequence
  outlined in the figure.
 }
\end{figure}

\begin{figure}[htb]
\centerline{\includegraphics[width=7.0in]{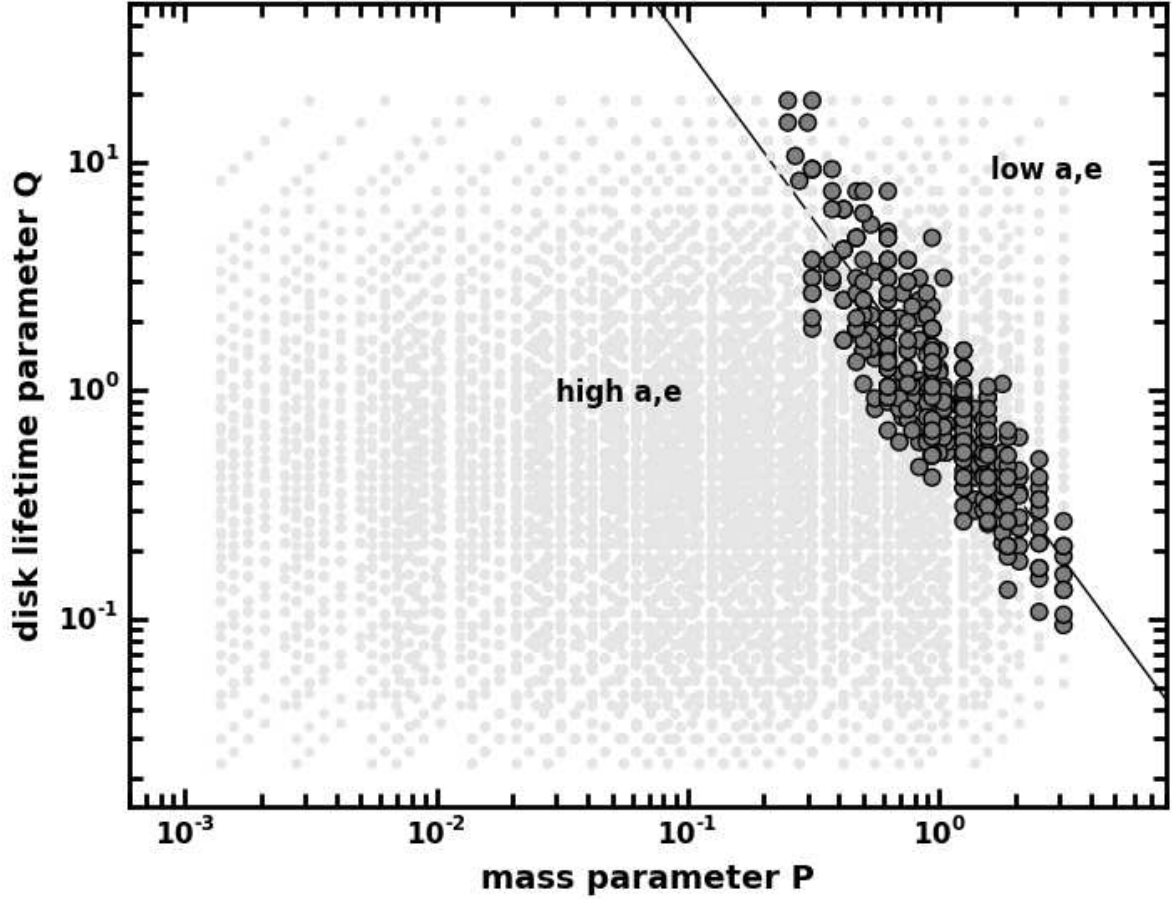}}
\caption{\label{fig:pixx} Disk lifetime and mass parameters describing
  outcomes for scattered planets in evolving disks.  Each point in
  this space of mass and disk lifetime parameters ($P$ and $Q$; see
  Equations~(\ref{eq:P}) and (\ref{eq:Q})) corresponds to an
  individual simulation with a unique set of planet and disk
  configurations. Dark circles with black outlines indicate successful
  models, which roughly match the orbital parameters of \PlanetX\ in
  \citet{batygin2016} and are located in the shaded region in
  Figure~\ref{fig:aep}. The line running through these points is from
  the approximation in Equation~(\ref{eq:PQ}).  The light gray points
  points, with the lightest shade of gray, cover the unsuccessful
  models where orbits are either too remote and eccentric or too close
  to the Sun and circular. }
\end{figure}

\textit{In situ} formation of \PlanetX\ is possible when disk evolution
produces a massive ring of solids beyond 100~AU.  Coagulation may then
grow super-Earths in 1--5~Gyr out to distances of 750~AU 
\citep{kb2015a, kb2016b}. In this mechanism, super-Earths reside on
fairly circular orbits.  For comparison, scattered planets can damp 
to circular orbits only inside of $\sim 200$~AU (see Fig.~\ref{fig:aep}). 
Thus, a \PlanetX\ found at a large orbital distance with low eccentricity
and low inclination strongly favors \textit{in situ} formation.  While 
not the favored choice of \citet{batygin2016}, it is unclear whether 
current observations explicitly rule out circular orbits for the massive 
perturber. 

Other purely dynamical events can also produce a \PlanetX. In the Nice
model \citep{tsig2005}, a dynamical instability after the gaseous disk
has dispersed can scatter a fully-formed giant planet into the outer
solar system. Most scattered planets are ejected \citep[e.g.,][]{nesvorny2011}.
If damping within a residual, low surface density gaseous disk is possible,
some scattered planets might be retained on high eccentricity ($e\gtrsim 0.9$),
low inclination ($i \lesssim 10^\circ$) orbits
\citep[e.g.,][]{marzari2010, raymond2010}. For a massive \PlanetX, the
high $e$ orbit might distinguish this mechanism from our model, where
scattering occurs when the inner edge of the disk lies much closer to
the Sun.

A passing star --- perhaps a member of the Sun's birth cluster
\citep{adams2001} --- can also relocate \PlanetX\ in the outer solar
system.  Outcomes vary widely, depending on the planet's initial orbit
\citep[e.g.][]{koba2001, kb2004d, morby2004a, brasser2006, kaib2008}. If the
planet starts on a circular orbit in the ecliptic plane, a stellar
flyby will give it a strong kick in eccentricity but only a mild boost
in perihelion distance and inclination. Thus, if the planet's
present-day semimajor axis is above 400~AU, it is likely to have an
eccentricity of 0.9 or more, unless it formed well beyond Neptune.


A passing star can yield a broader range of outcomes if \PlanetX\ were
on an eccentric orbit at the time of the flyby, perhaps as a result of
a previous stellar encounter or the scattering mechanism considered
here. Alternatively, if the Sun captured \PlanetX\ from the passing
star, the possibilities are even greater \citep[e.g., Figures 2 and 3
of][see also \citealt{morby2004a}, \citealt{levison2010b},
  \citealt{jilkova2015}]{kb2004d}. However, the likelihood of this
eventuality seems low \citep{li2016}.

Finally, we consider the effect of tides from the Galactic environment. 
For Oort cloud comets, the gravitational potential of the Galaxy 
dominates the orbital evolution \citep{heisler1986, duncan1987}.  
However, tidal effects become weak inside $10^4$~AU; evolutionary time 
scales are then long, 100~Myr or more. For objects with a semimajor axis 
within 1000~AU, the Galactic tide causes only small changes in the orbit
over the age of the solar system \citep{higuchi2007, brasser2008}.  In 
our static disk models, the semimajor axes we consider are at 800~AU and 
smaller; a putative \PlanetX\ is then shielded from tidal interaction.  
In the evolving disk models, we use a maximum semimajor axis of 1800~AU.  
However, in successful models, the semimajor axis falls well below 700~AU 
within 10~Myr, well within the tidal evolution time scale.

In the absence of any gas, the Galactic tide can influence the orbit
of a \PlanetX\ initially scattered beyond $\sim 1000$~AU within a 
billion years of the solar system's formation.  Torque from the 
Galactic potential then raises both the perihelion and the inclination
of the orbit \citep[e.g.,][]{duncan1987, higuchi2007, brasser2008}. 
The hallmark of this process would be a semimajor axis exceeding 1000~AU, 
a perihelion distance of at least 100~AU, an eccentricity of 0.8--0.9, 
and an inclination that may be anywhere from $i = 0^\circ$ to $\sim$135$^\circ$ 
\citep[e.g., Figures~9 and 11 of][]{higuchi2007}.

Tides from the Sun's birth cluster may have had an even more dramatic
effect than the Galactic tide \citep[e.g.,][]{brasser2006}. However, if
this cluster was typical of other embedded clusters, it would have
disintegrated quickly, within 2--3~Myr \citep[see][]{lada2003}. The
density of stars in the cluster, the Sun's orbit through it, and the
timing of the cluster dispersal relative to the formation of the gas
giants are all uncertain.  If \PlanetX's final orbit was determined by
interactions during this phase of the Sun's history, then its high
perihelion distance would also likely be accompanied by a high
inclination \citep[e.g., Figures 6 and 8 of][]{brasser2006}.

Observations of exoplanetary systems provide ways to test these scenarios.
Over the next 10--20~yr, direct imaging will probably yield large samples 
of gas giants at large $a$. Comparison of the observed properties of these 
systems with the predictions of numerical simulations should enable 
constraints on the likelihood of any particular theoretical model. 
For stars with ages of 5--10~Myr, current data suggest many systems with 
$\lesssim$ 1 Jupiter mass of gas \citep[e.g.,][]{dent2013}. Expanding 
surveys to older stars and reducing upper limits on the mass in gas by 
an order of magnitude would challenge some of our scattering models.

In the solar system, identifying \PlanetX\ and new dwarf planets is
essential for making progress.  As outlined in \citet{batygin2016},
larger samples of dwarf planets provide additional constraints on any
\PlanetX. A robust detection of a massive perturber
\citep[see][]{cowan2016, linder2016, ginzburg2016, delafuenta2016} and
direct measurement of orbital elements allow discrimination between
the various possibilities for the origin and evolution of \PlanetX.
If interactions with a gas disk turn out to be important, the next
step is to obtain more realistic predictions of scattering outcomes
with hydrodynamical simulations.  Combined with observations of
exoplanets, these advances might determine the fate of scattered
planets.

\acknowledgements

We are grateful to M.\ Geller, J. Najita and D. Wilner for comments and 
helpful discussions.  NASA provided essential support for this program 
through a generous allotment of computer time on the NCCS 'discover' 
cluster and {\it Outer Planets Program} grant NNX11AM37G.

\bibliography{planets}{}
\bibliographystyle{apj}

\begin{table}
\caption{Disk Parameters}\label{tab:parmsdisk}
\begin{tabular}{lcll}
\multicolumn{4}{c}{\ }
\\
\hline
\ \ \ \ \ \ Name & \ \ \ Symbol \ \ \ & Value or Range & Units  \\
\hline
radial length scale & $a_0$ & 1 & AU 
\\
scale height factor & $\hoh$ & 0.05 & --
\\[5pt]
\multicolumn{4}{l}{\underline{\it static disk}}
\\
surface density & $\Soh$ & 2, 10, 20, 50 & g/cm$^2$ 
\\
initial inner edge & $\ain$ & 50, 100, 200 & AU
\\
outer edge & $\aout$  & 1600 & AU 
\\[5pt]
\multicolumn{4}{l}{\underline{\it evolving disk}}
\\
surface density & $\Soh$ & 50, 100, 200, 500, 1000 & g/cm$^2$ 
\\
initial inner edge & $\ain$ & 20, 60, 100, 140, 180 & AU
\\
outer edge & $\aout$  & 800 & AU 
\\
opening rate ($\aindot$) & $\openrate$  & 20, 40, 60, 80 & AU/Myr 
\\
decay time & $\tau$  & 2, 4, $\infty$ & Myr 
\\
\hline
\end{tabular}
\end{table}

\begin{table}
\caption{Planet Parameters}\label{tab:parmsplanet}
\begin{tabular}{lcll}
\multicolumn{4}{c}{\ }
\\
\hline
\ \ \ \ \ \ Name & \ \ \ Symbol \ \ \ & Value or Range & Units  
\\
\hline
mass & $\mp$ & 1, 5, 10, 15, 20, 30, 50 & $\Mearth$
\\
mean density & $\rhop$ & 1.33 & g/cm$^3$
\\
initial perihelion & $\peri$ & 10 & AU
\\
initial aphelion & $\apo$ & 1600, 2000, 2400, ..., 3600  & AU
\\
inclination & $i$ & 0 & rad
\\
\hline
\end{tabular}
\end{table}

\newpage

\end{document}